\documentclass[a4paper,fleqn]{cas-sc}

\usepackage[authoryear,longnamesfirst]{natbib}

\usepackage{amsmath}
\usepackage{amssymb}
\usepackage{graphicx}
\usepackage{booktabs}

\usepackage{lastpage}
\AtBeginDocument{}

\newcommand{\dEperp}{\delta E_{\perp}}
\newcommand{\dBperp}{\delta B_{\perp}}
\newcommand{\dEpar}{\delta E_{\parallel}}
\newcommand{\vA}{v_{\rm A}}
\newcommand{\kperp}{k_{\perp}}
\newcommand{\rhoi}{\rho_i}
\newcommand{\rhoe}{\rho_e}

\newcommand{\vbetai}         {14.4\xspace}

\newcommand{\vrhoiOverdi}    {3.79\xspace}      
     
\newcommand{\vVflow}         {159\xspace}       
\newcommand{\vVflowErr}      {33\xspace}

\newcommand{\valphaFluid}    {-1.81\xspace}     
\newcommand{\valphaFluidErr} {0.08\xspace}      
\newcommand{\vfbreak}        {0.48\xspace}      
\newcommand{\vfbreakErr}     {0.06\xspace}      
\newcommand{\vkdiBreak}      {0.85\xspace}      
\newcommand{\vkdiBreakErr}   {0.11\xspace}      
\newcommand{\vkrhoiBreak}    {3.23\xspace}      
\newcommand{\vkrhoiBreakErr} {0.42\xspace}

\newcommand{\valphaKin}      {-2.86\xspace}     
\newcommand{\valphaKinStat}  {0.06\xspace}      
\newcommand{\valphaKinWin}   {0.18\xspace}      
\newcommand{\valphaKinComb}  {0.19\xspace}      
\newcommand{\valphaKinFGM}   {-2.90\xspace}

\newcommand{\vfgmScmLo}      {1.4\xspace}       
\newcommand{\vfgmScmHi}      {17.4\xspace}      
\newcommand{\valphaPerp}     {-2.85\xspace}     
\newcommand{\valphaPerpErr}  {0.06\xspace}      
\newcommand{\valphaPar}      {-2.89\xspace}     
\newcommand{\valphaParErr}   {0.09\xspace}      
\newcommand{\vCB}            {0.310\xspace}     
\newcommand{\vCBscm}         {0.332\xspace}     
\newcommand{\vPratio}        {2.22\xspace}      
\newcommand{\vPratioscm}     {2.01\xspace}

\newcommand{\vpeakEpar}      {2.05\xspace}      
\newcommand{\vEBmeas}        {2.41\xspace}      
\newcommand{\vEBadv}         {1.84\xspace}      
\newcommand{\vEBnonideal}    {1.29\xspace}      
\newcommand{\vEBslope}       {+0.92\xspace}

\newcommand{\vtwoFluidRatio} {0.68\xspace}      
\newcommand{\vtwoFluidOffset}{32\xspace}        
\newcommand{\vfrozenRatio}   {0.74\xspace}      
        
\newcommand{\vpressureR}     {-0.64\xspace}     
\newcommand{\vpressureSlope} {-0.81\xspace}     
\newcommand{\vcompressR}     {-0.30\xspace}     
\newcommand{\vanisoMed}      {0.978\xspace}     
\newcommand{\vbetaParMed}    {13.5\xspace}      
\newcommand{\vmirrorThresh}  {1.106\xspace}

\newcommand{\vrmsBqLo}       {3.6\xspace}       
\newcommand{\vrmsBqHi}       {6.1\xspace}       
\newcommand{\vslamsMax}      {2.36\xspace}      
\newcommand{\vdBcompress}    {0.34\xspace}      
\newcommand{\vthetakB}       {75\xspace}

\newcommand{\vqualityPass}   {21\xspace}        
\newcommand{\vkdotV}         {0.81\xspace}      
\newcommand{\vkdotVloIQR}    {0.77\xspace}      
\newcommand{\vkdotVhiIQR}    {0.87\xspace}

\newcommand{\vdBoverBtot}    {0.54\xspace}      
\newcommand{\vdBoverBcomp}   {0.46\xspace}      
\newcommand{\vdBoverBband}   {0.40\xspace}      
\newcommand{\voctFluid}      {0.35\xspace}      
\newcommand{\voctBreak}      {0.14\xspace}      
\newcommand{\voctKin}        {0.04\xspace}      
\newcommand{\voctTen}        {0.01\xspace}

\newcommand{\vthetaWinLo}    {65.7\xspace}      
\newcommand{\vthetaWinHi}    {75.0\xspace}

\usepackage{xspace}

\begin{document}
\let\WriteBookmarks\relax
\renewcommand{\textfraction}{0.05}
\renewcommand{\topfraction}{0.95}
\renewcommand{\bottomfraction}{0.95}
\renewcommand{\floatpagefraction}{0.75}
\setcounter{topnumber}{3}
\setcounter{bottomnumber}{2}
\setcounter{totalnumber}{4}

\shorttitle{Kinetic Alfv\'{e}n wave turbulence at ion scales in the magnetosheath}
\shortauthors{Mani K Chettri et al.}

\title[mode = title]{Electromagnetic Signatures of Kinetic Alfv\'{e}n Wave Turbulence at Ion Inertial Scales in Earth's High-$\beta$ Magnetosheath}

\author[1]{Mani K Chettri}[orcid=0009-0000-1368-9263]
\fnmark[1]
\credit{Conceptualization, Data curation, Software, Formal analysis,
Writing -- original draft}

\author[1]{Rupak Mukherjee}[orcid=0000-0003-3955-7116]
\credit{Supervision, Writing -- review \& editing}

\author[2]{Hemam D.\ Singh}[orcid=0009-0003-3061-8944]
\cormark[1]
\ead{hemam.singh@nsut.ac.in}
\credit{Conceptualization, Supervision, Writing -- review \& editing}

\affiliation[1]{organization={Department of Physics, Sikkim University},
            city={Gangtok},
            postcode={737102},
            state={Sikkim},
            country={India}}

\affiliation[2]{organization={Department of Physics,
            Netaji Subhas University of Technology},
            city={New Delhi},
            postcode={110078},
            country={India}}

\cortext[1]{Corresponding author}
\fntext[1]{Electronic email: mkchettri8@gmail.com}

\begin{abstract}
We present a multi-diagnostic electromagnetic analysis of a dayside
magnetosheath interval observed by the Magnetospheric Multiscale (MMS) mission
on 2015 December 28 at high ion beta ($\beta_i = \vbetai$), using burst-mode
measurements. We combine the scale-dependent electric-to-magnetic field ratio,
the parallel electric field, the magnetic compressibility, the wave-normal
angle and circular-polarization measure, and the ion--electron frozen-in
residuals. The electric-to-magnetic field ratio rises with wavenumber with a
log--log slope of \vEBslope, whereas the ideal advection contribution remains
flat, supporting dispersive fluctuations. A wave-normal angle of
$\vthetakB^\circ$ and a circular-polarization measure centred near zero support
quasi-perpendicular propagation, while the compressibility and the ion
temperature anisotropy do not support mirror-mode dominance. A continuous
broken power-law fit gives $f_{\rm break} = \vfbreak \pm \vfbreakErr$~Hz,
corresponding to $\kperp d_i = \vkdiBreak \pm \vkdiBreakErr$ and
$\kperp\rhoi = \vkrhoiBreak \pm \vkrhoiBreakErr$. Because $d_i$ and $\rhoi$ are
well separated at this $\beta_i$, the break is more consistent with the ion
inertial scale than with the ion gyroscale. Four-spacecraft phase differencing
independently supports the Doppler conversion around the break. Above the
break, the kinetic-range index is $\valphaKin \pm \valphaKinStat$~(stat)
$\pm \valphaKinWin$~(fit window), closer to the weakly damped dispersive limit
than to the strongly damped limit, with a fluid-scale index of
$\valphaFluid \pm \valphaFluidErr$ below it. The observations are consistent
with quasi-perpendicular KAW-like turbulence, but do not uniquely identify the
mode; the spectral index constrains the cascade rather than the dissipation
mechanism.
\end{abstract}


\maketitle

\section{Introduction}
\label{sec:intro}

Earth's magnetosheath, the turbulent compressed plasma occupying
the region immediately downstream of the bow shock, provides an
accessible high-$\beta$ laboratory for studying kinetic-scale plasma
turbulence and energy dissipation.
Unlike the ambient solar wind, where $\beta \sim 1$ is typical, the
magnetosheath regularly sustains $\beta \sim 1$--$10$
\citep{lucek05,alexandrova08,macek18,sahraoui20}, placing it in a regime where
the dispersive and dissipative properties of kinetic Alfv\'{e}n waves
(KAWs) are strongly $\beta$-dependent and wave-particle resonances
are particularly efficient.

KAWs constitute the dispersive continuation of the shear Alfv\'{e}n
mode at perpendicular scales $\kperp\rhoi \gtrsim 1$, where $\rhoi$
is the ion thermal gyroradius.
Two electromagnetic signatures distinguish them from ideal
magnetohydrodynamic (MHD) Alfv\'{e}n waves: a nonzero parallel
electric field $\delta E_{\parallel}$, which drives field-aligned
electron acceleration, and an enhanced ratio of perpendicular electric
to magnetic field fluctuations $\dEperp / (\dBperp \vA) > 1$
\citep{hasegawa76,hollweg99,stasiewicz00}.
The parallel electric field is particularly important because it
enables Landau resonance with electrons satisfying
$v_{\parallel} \approx \omega / k_{\parallel}$, establishing KAWs as
primary vectors of turbulent energy dissipation in collisionless
high-$\beta$ plasmas \citep{howes14,chen19,afshari21,schekochihin09}.

The Magnetospheric Multiscale (MMS) mission, with its burst-mode
sampling of 128~Hz for magnetic fields and concurrent electric field
and particle measurements, has substantially advanced observational
access to these kinetic-scale processes
\citep{burch16,russell16,pollock16}. Studies using MMS have reported individual KAW wave packets in
magnetosheath reconnection regions \citep{gershman17,stawarz22},
field-particle correlation evidence of Landau damping \citep{afshari21},
and systematic spectral steepening near
the ion gyrofrequency \citep{macek18,macek23}. KAW turbulence and nonlinear wave coupling have similarly been
reported in the inner heliosphere and the plasma sheet boundary
layer \citep{chettri2024nonlinear,chettri2026mms}, motivating the
multi-diagnostic approach applied here.

Prior spectral analyses of the 2015 December 28 MMS magnetosheath
interval by \citet{macek18} and \citet{macek23} established the
two-range spectral structure and documented ion-scale steepening at
conditions of $\beta \sim 2$--$5$.
The present study extends that work in three respects.
First, we analyze this interval at the event-specific ion beta
$\beta_i = 14.4$, which places it at the high-$\beta$ end of
magnetosheath conditions and allows the ion inertial length $d_i$
and the ion gyroradius $\rho_i$ to be distinguished observationally:
the two scales differ here by a factor of \vrhoiOverdi{}. This separation allows us to
test whether the measured break is more consistent with $d_i$ or with
$\rho_i$. At $\beta_i \gg 1$ the dispersive modification to Alfv\'{e}n waves is
expected to set in at $\kperp d_i \sim 1$ \citep{hasegawa76,stasiewicz00}. The same test is far harder at
$\beta \sim 1$--$5$, where the two scales differ by at most a factor of two.
{}
Second, we combine the scale-dependent electric-to-magnetic field ratio,
the parallel electric field, the magnetic compressibility, the wave-normal
angle, the circular-polarization measure, and the ion and electron frozen-in
residuals.
None of these measurements identifies a KAW uniquely, but together they
constrain the principal alternative interpretations and allow the fluctuations
to be assessed for consistency with KAW-like turbulence. The comparison of the
ion and electron frozen-in residuals provides a direct constraint on Hall-scale
decoupling, since the decoupling of the ions from the field at $\kperp d_i \sim 1$ is
the Hall physics that makes a kinetic Alfv\'{e}n wave dispersive.
Third, we place the kinetic-range spectral index within the
theoretical framework of \citet{boldyrev12} and \citet{passot15} to
obtain a direct observational constraint on the {}state of the cascade, and discuss how this approach could be extended to statistical
surveys using standard burst-mode magnetometer data.
We are explicit that a spectral index constrains the cascade and not the
dissipation mechanism.

\section{Data and Interval Selection}
\label{sec:data}

We analyze MMS1 Level-2 burst-mode data from 2015 December~28,
01:48:00--01:52:30~UT, during which the spacecraft was positioned in
the dayside magnetosheath at $(X, Y, Z)_{\rm GSE} = (10.2, -5.8,
-1.0)~R_E$. The interval is shown in Figure~\ref{fig:mms_panels}. The
displayed interval spans 270~s; the common grid on which the joint
multi-instrument quantities are computed begins at 01:48:04~UT and spans
266~s. This interval was identified in prior MMS analyses as exhibiting a
well-defined two-range magnetic power spectrum with steepening near
ion scales and pronounced Alfv\'{e}nic fluctuations
\citep{macek18,macek23}, confirming its suitability as a testbed for
kinetic-scale electromagnetic analysis. Those studies focused on the magnetic field spectrum; the plasma
$\beta$ environment for this specific event was not their primary
reported quantity.

The present study independently characterizes the plasma environment
using burst-mode FPI moments computed directly over the 4.5-minute
window, yielding $\beta_i = 14.4$, $\beta_e = 1.6$ and
$\beta_{\rm tot} = 16.0$ (Table~\ref{tab:plasma_params}).
Fast-survey FPI data over the full 70-minute magnetosheath pass
confirm that $\beta_i$ remains elevated ($\beta_i \sim 10$--$40$)
throughout, with a median of $\approx 14.8$ during the burst interval,
{}so the analyzed
interval is not an isolated transient excursion.

{}Four MMS instrument suites are used. The FluxGate Magnetometer (FGM) provides vector magnetic field
measurements at 128~Hz \citep{russell16}.
The Search-Coil Magnetometer (SCM) provides the magnetic fluctuation
vector at 8192~samples~s$^{-1}$ in burst mode (\texttt{scb} product;
\citealt{lecontel16}), which we use to establish the frequency range over which
FGM is reliable in this interval and to refit the kinetic-range index.
The Electric Double Probe (EDP) provides three-component electric
field measurements at comparable cadence \citep{torbert16}.
The Fast Plasma Investigation (FPI) provides ion and electron moments
at 150~ms and 30~ms cadence respectively, which we use for background
plasma characterization and, in Section~\ref{sec:discussion}, for the
frozen-in test \citep{pollock16}.
The four spacecraft were separated by 33--41~km during this interval, that
is $0.73$--$0.92\,d_i$, so the tetrahedron resolves wavenumbers up to the
spatial Nyquist limit $k_{\max} = \pi/d_{\max}$, corresponding to
$\kperp d_i = 3.4$ or 1.93~Hz under Taylor's hypothesis
(Figure~\ref{fig:multisc}a). Spacecraft positions are taken from the MEC
ephemeris.
Data are retrieved and processed using the PySPEDAS library
\citep{angelopoulos19,grimes22}, including standard spike removal,
coordinate transformation to Geocentric Solar Ecliptic (GSE)
coordinates, and linear detrending prior to spectral analysis.
For wave analysis, a background field is estimated with a $10$~s sliding
  window (corresponding to the lower cutoff of the $[0.1,\,2.0]$~Hz
  bandpass) and subtracted; electric and magnetic fluctuations are then
decomposed into field-aligned (parallel) and perpendicular components. Throughout this paper, `parallel' and `perpendicular' refer to
directions relative to the background magnetic field $\mathbf{B}_0$.
A bandpass filter of $[0.1, 2.0]$~Hz isolates fluctuations in the
transition from {}fluid to kinetic scales.
Power spectral densities are computed using Welch's method
\citep{welch67} with a Hanning window{}, applied to the trace of
the spectral matrix rather than to a single component, so that the result does
not depend on the choice of coordinate system. Spectra are binned
logarithmically at 20 bins per decade before fitting, because linearly spaced
bins place about three times as many points between 10 and 20~Hz as between 3
and 6~Hz and therefore weight a least-squares fit towards the upper end of the
range.

\begin{figure}[pos=htbp]
\centering
\includegraphics[width=\columnwidth]{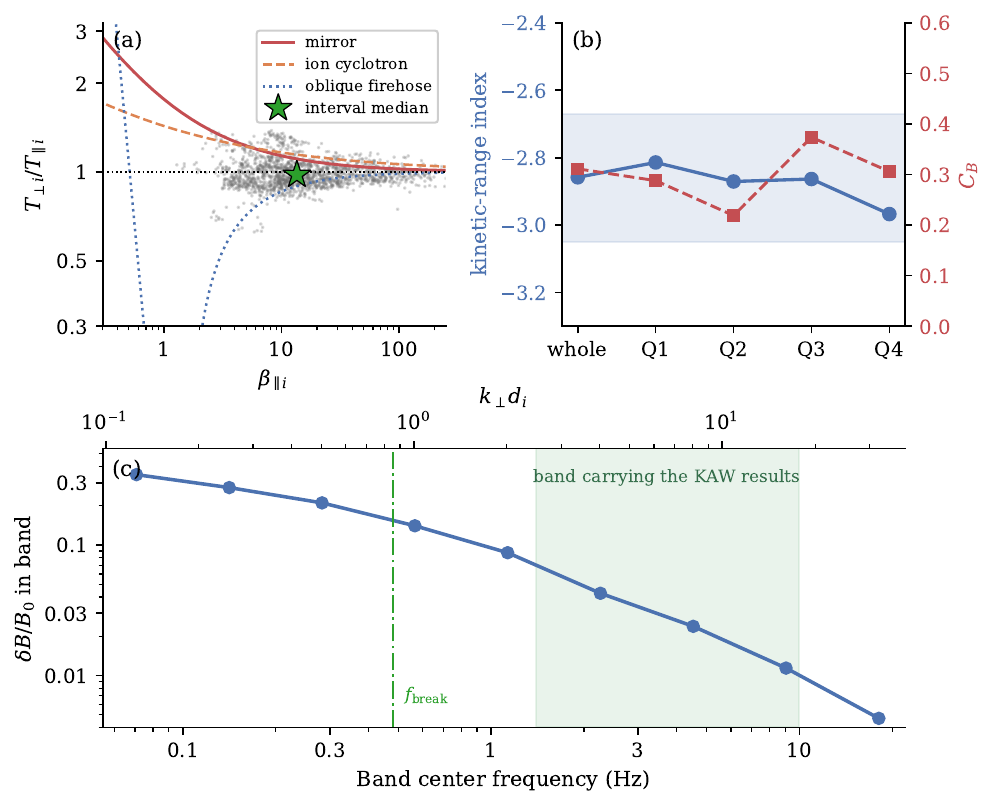}
\caption{Interval quality and the scale dependence of the field-aligned
decomposition. (a)~Ion temperature anisotropy against
$\beta_{\parallel i}$ for individual 150~ms FPI samples (gray), with the mirror,
ion-cyclotron and oblique-firehose thresholds of \citet{hellinger06} for a
growth rate of $10^{-3}\Omega_{ci}$. The star marks the interval median,
$T_{\perp i}/T_{\parallel i} = \vanisoMed$ at $\beta_{\parallel i} = \vbetaParMed$, which
lies below the mirror threshold of \vmirrorThresh at that $\beta$. (b)~Kinetic-range
spectral index (blue, left axis) and magnetic compressibility $C_B$ (red, right
axis) computed independently in each 66~s quarter. The shaded band is the
$\pm\valphaKinComb$ combined uncertainty on the whole-interval index quoted in
Section~\ref{sec:gauge}; the quarter-to-quarter spread of the index, 0.16, is
smaller than it.
(c)~$\delta B/B_0$ in successive octave bands, with the corresponding
$\kperp d_i$ on the upper axis. This ratio determines whether a
field-aligned decomposition is justified at a given scale: it falls from
0.35 at $\kperp d_i = 0.13$, through 0.14 across the break (dash-dotted line),
to 0.04 above 1.6~Hz and 0.01 by 10~Hz. The decomposition therefore carries a
real ambiguity below the break and is well founded across the shaded band,
which is where the results of Sections~\ref{sec:identification}
and~\ref{sec:polarization} are measured.}
\label{fig:interval}
\end{figure}

The magnetic fluctuation amplitude relative to the background field is
$|\delta\mathbf{B}|/\langle|B|\rangle = \vdBoverBtot$ over the interval as a
whole, measured about a 10~s mean; the compressive part alone gives
$\delta|B|/\langle|B|\rangle = \vdBoverBcomp$ about the interval mean, and the
value within the $[0.1, 2.0]$~Hz band is $\vdBoverBband$. The integrated
fluctuation level is therefore not small. The ratio is, however, strongly scale
dependent: in octave bands it decreases from \voctFluid at fluid scales
($\kperp d_i = 0.13$), to \voctBreak near the spectral break, to \voctKin above
1.6~Hz and to \voctTen by 10~Hz (Figure~\ref{fig:interval}c). The field-aligned
decomposition is therefore less certain below the break, but is much better
justified across the kinetic range in which the results of
Sections~\ref{sec:identification} and~\ref{sec:polarization} are obtained.

The time-averaged plasma parameters computed directly from the
burst-mode data are listed in Table~\ref{tab:plasma_params}.
The background magnetic field is $B_0 = 17.8 \pm 8.2$~nT, the ion
density $n_i = 25.6 \pm 7.6$~cm$^{-3}$, and the ion (electron)
temperature $T_i = 0.44$~keV ($T_e = 0.047$~keV), giving an ion beta
$\beta_i = 14.4$.
The Alfv\'{e}n speed is $\vA = 77$~km~s$^{-1}$, and the ion inertial
length $d_i = 45.0$~km.
The ion thermal gyroradius is $\rhoi = 170.4$~km, with
$\rhoi / d_i = \sqrt{\beta_i} = \vrhoiOverdi$ as expected.
The bulk plasma flow velocity $V_{\rm flow} = 159 \pm 33$~km~s$^{-1}$,
giving $V_{\rm flow} / \vA = 2.1$; this satisfies the Taylor frozen-in
criterion $V_{\rm flow} > \vA$ marginally, and we assess its
implications in Section~\ref{sec:tworange}.

Because shock-generated and compressive structures alter spectral slopes,
we have examined the interval for them directly. The median ion temperature
anisotropy is $T_{\perp i}/T_{\parallel i} = \vanisoMed$ (5--95~\% range
0.864--1.233), slightly below unity, so the median state does not have the
perpendicular ion-temperature anisotropy required to drive the proton mirror
instability. Using the mirror-instability threshold fit of \citet{hellinger06},
the threshold at the median $\beta_{\parallel i} = \vbetaParMed$ is
$T_{\perp}/T_{\parallel} = \vmirrorThresh$, above the measured median value.
Individual 150~ms samples cross this threshold 13.9~\% of the time. Restricting
the calculation to $0.5 < |B|/\langle|B|\rangle < 1.5$ gives a similar fraction
of 14.6~\%, showing that these crossings are not caused by low-field
measurements. The interval therefore lies close to marginal stability, while
its median plasma state remains below the mirror threshold. Taken
together with the compressibility reported in
Section~\ref{sec:polarization} and the weak pressure-balanced component
described next, these results do not support a mirror-mode-dominated
interval.

We have also characterized the compressive structures that are present.
The correlation between the $|B|$ and $n_i$ fluctuations about a 10~s running
mean is $r = \vcompressR$ ($N = 1641$), compared with $r = -0.6$ to $-0.9$ for mirror
modes and pressure-balanced structures and strongly positive values for
fast-mode compressions. A weak pressure-balanced component is therefore
present, but it does not dominate the interval. We then searched for
large-amplitude magnetic structures. SLAMS are conventionally identified as
field enhancements of two to four times the background lasting several seconds;
here $|B|$ never exceeds \vslamsMax times the 10~s background and remains above
twice the background for only 0.81~\% of the interval, approximately 2~s in
total, in isolated spikes. We therefore find no sustained SLAMS-like or
shocklet-like magnetic enhancements in the field-magnitude time series. The rms
relative field-magnitude fluctuation about the 10~s background is
$\delta|B|/|B_0| = \vdBcompress$, typical of the magnetosheath.
Figure~\ref{fig:interval}(a) shows the anisotropy against
$\beta_{\parallel i}$ with the instability thresholds.

\begin{table}[width=\linewidth,pos=ht]
\caption{Plasma parameters for MMS1, 2015 December~28,
01:48:00--01:52:30~UTC (dayside magnetosheath).
Ion scalar temperature $T_i = (T_{i,\parallel} + 2T_{i,\perp})/3$
is used for $\beta_i$; perpendicular temperature $T_{i,\perp}$ is
used for $\rhoi$.
Wave amplitudes use a $[0.1, 2.0]$~Hz bandpass.
The three normalized $\delta E/\delta B$ entries are medians over the band.
The dispersive part is the non-ideal
remainder, and it is that value, not the measured ratio, that is used in the
identification of Section~\ref{sec:identification}.}
\label{tab:plasma_params}
\begin{tabular*}{\tblwidth}{@{} LLCC @{}}
\toprule
Parameter & Symbol & Value & Unit \\
\midrule
\multicolumn{4}{@{}l}{\textit{Observational parameters}} \\
SC position (GSE)              & $(X,Y,Z)$           & $(10.2,\ {-5.8},\ {-1.0})$ & $R_E$ \\
Background magnetic field      & $B_0$               & $17.8\pm8.2$     & nT \\
Ion number density             & $n_i$               & $25.6\pm7.6$     & cm$^{-3}$ \\
Ion temperature (scalar)       & $T_i$               & $0.442\pm0.120$  & keV \\
Ion temperature (perp)         & $T_{i,\perp}$       & $0.440\pm0.120$  & keV \\
Electron temperature           & $T_e$               & $0.047\pm0.004$  & keV \\
Temperature ratio              & $T_i/T_e$           & $9.3$            & -- \\
Ion beta                       & $\beta_i$           & $14.4$           & -- \\
Electron beta                  & $\beta_e$           & $1.6$            & -- \\
Total beta                     & $\beta_{\rm tot}$   & $16.0$           & -- \\
Bulk flow speed                & $V_{\rm flow}$      & $159\pm33$       & km~s$^{-1}$ \\
\midrule
\multicolumn{4}{@{}l}{\textit{Derived characteristic scales}} \\
Alfv\'{e}n speed               & $\vA$               & $77$             & km~s$^{-1}$ \\
Ion gyroradius ($\perp$)       & $\rhoi$             & $170.4$          & km \\
Ion inertial length            & $d_i$               & $45.0$           & km \\
Ion length ratio $(=\sqrt{\beta_i})$ & $\rhoi/d_i$         & $\vrhoiOverdi$ & -- \\
Ion cyclotron frequency        & $f_{ci}$            & $0.271$          & Hz \\
\midrule
\multicolumn{4}{@{}l}{\textit{Wave properties (bandpass $[0.1, 2.0]$~Hz)}} \\
RMS electric field       & $\langle\delta E_\perp^2\rangle^{1/2}$ & $1.08$ & mV~m$^{-1}$ \\
RMS magnetic field       & $\langle\delta B_\perp^2\rangle^{1/2}$ & {}$5.58$ & nT \\
Peak parallel field      & $E_{\parallel,\rm max}$ & {}$\vpeakEpar$             & mV~m$^{-1}$ \\
{}
Norm.\ $\delta E/\delta B$, measured & $\dEperp/(\vA\dBperp)$ & $\vEBmeas$ & -- \\
\quad advection contribution & $|{-}\mathbf{V}\times\mathbf{B}|_\perp/(\vA\dBperp)$ & $\vEBadv$ & -- \\
\quad non-ideal remainder & $|\mathbf{E}{+}\mathbf{V}\times\mathbf{B}|_\perp/(\vA\dBperp)$ & $\vEBnonideal$ & -- \\
\bottomrule
\end{tabular*}
\end{table}

\section{{}Multi-Diagnostic
Assessment of KAW-Like Fluctuations}
\label{sec:identification}

\begin{figure}[pos=htbp]
\centering
\includegraphics[width=0.88\columnwidth]{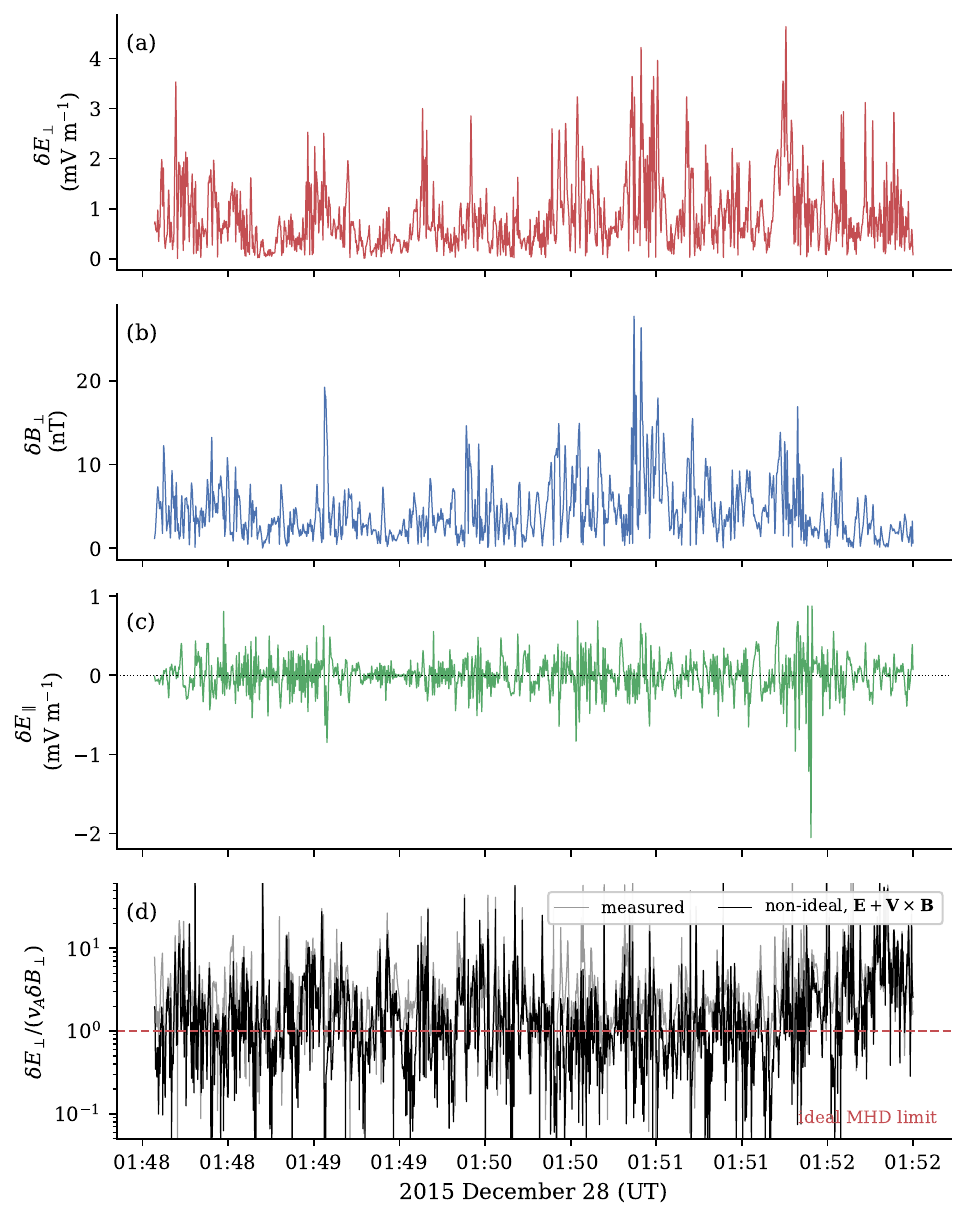}
\caption{MMS1 electromagnetic field analysis, 2015 December~28,
01:48:00--01:52:30~UT ($\beta_i = 14.4$, $V_{\rm flow} = 159$~km~s$^{-1}$).
(a) Amplitude of perpendicular electric field fluctuations
$\dEperp$ (RMS $1.08$~mV~m$^{-1}$, peak $4.5$~mV~m$^{-1}$).
(b) Amplitude of perpendicular magnetic field fluctuations
$\dBperp$ (RMS {}$5.58$~nT, peak $\sim27$~nT).
(c) Amplitude of parallel electric field $\dEpar$ (peak {}$\vpeakEpar$~mV~m$^{-1}$,
from the EDP field-aligned product).
(d) Normalized ratio $\dEperp / (\dBperp \vA)$ {}as
measured (gray) and after removal of the ideal advection field
$-\mathbf{V}\times\mathbf{B}$ (black). The advection contribution alone has a
median of $\vEBadv$; the non-ideal remainder, which is the quantity that carries
information about dispersion, has a median of $\vEBnonideal$. The dashed line marks
the ideal MHD limit (ratio = 1).
A $[0.1, 2.0]$~Hz bandpass and 10~s background subtraction are applied.}
\label{fig:mms_panels}
\end{figure}

Figure~\ref{fig:mms_panels}, panel~(a) of which is adapted from
\citet{chettri2025damped}, presents the four-panel electromagnetic
analysis.
Panel~(a) shows the amplitude of perpendicular electric field
fluctuations $\dEperp$, which range up to $\sim4.5$~mV~m$^{-1}$.
Panel~(b) shows perpendicular magnetic field fluctuations $\dBperp$
(RMS {}$5.58$~nT), spanning up to $\sim27$~nT.
Both time series are correlated and display a distinctly intermittent,
bursty character rather than a smooth continuous oscillation,
consistent with the spatial inhomogeneity of turbulence in the
shocked magnetosheath \citep{roberts18}.

Panel~(c) shows the parallel electric field amplitude
$\dEpar$, which reaches up to {}$2.05$~mV~m$^{-1}$.
This quantity is the most direct electromagnetic discriminator between
KAWs and ideal MHD Alfv\'{e}n waves: MHD strictly requires
$\delta E_{\parallel} = 0$, while KAWs generically support finite
$\delta E_{\parallel}$ through the combined effects of electron
inertia and electron thermal pressure gradients along the field
\citep{lysak96,wygant02,ergun16}.
We quote $\dEpar$ from the EDP field-aligned data product, which gives a
peak value of \vpeakEpar~mV~m$^{-1}$. Projecting the full three-component field
onto $\hat{\mathbf{b}}$ instead gives a peak of $3.23$~mV~m$^{-1}$; the axial
double probe contributes half the variance of that projection and the two
estimates correlate at only $r = 0.41$, and we therefore use the field-aligned
product throughout.
{}
The axial component is measured by the axial double probe and is not
reconstructed under the assumption $\mathbf{E}\cdot\mathbf{B} = 0$. A finite
$\dEpar$ excludes ideal-MHD Alfv\'{e}n fluctuations, but it is not unique to
KAWs: oblique slow modes and other electron-pressure-supported fluctuations can
also have a finite parallel electric field.

Panel~(d) shows the normalized ratio
$\dEperp / (\dBperp \vA)$.
For ideal MHD Alfv\'{e}n waves this quantity equals unity; for KAWs
it exceeds unity by an amount that increases with $\kperp\rhoi$ and
depends on plasma $\beta$ \citep{stasiewicz00,salem12,bale05}.
{}
A ratio above unity does not on its own establish dispersion here. In
ideal MHD any structure carried past the spacecraft satisfies
$\mathbf{E} = -\mathbf{V}\times\mathbf{B}$, and this interval has
$V_{\rm flow}/\vA = 2.1$, so advection alone produces a ratio well above unity.
Separating the two contributions, the measured median of \vEBmeas contains
\vEBadv from the advection field $-\mathbf{V}\times\mathbf{B}$; the measured
$\delta\mathbf{E}$ correlates with that field at $r = 0.85$ with a regression
slope of $0.97$, and advection accounts for 68~\% of the perpendicular electric
field variance. The dispersive part is the
remainder,
$|(\mathbf{E}+\mathbf{V}\times\mathbf{B})_{\perp}|/(\dBperp \vA) = \vEBnonideal$,
which we quote in place of the uncorrected ratio. Panel~(d) shows both.

\begin{figure}[pos=htbp]
\centering
\includegraphics[width=\columnwidth]{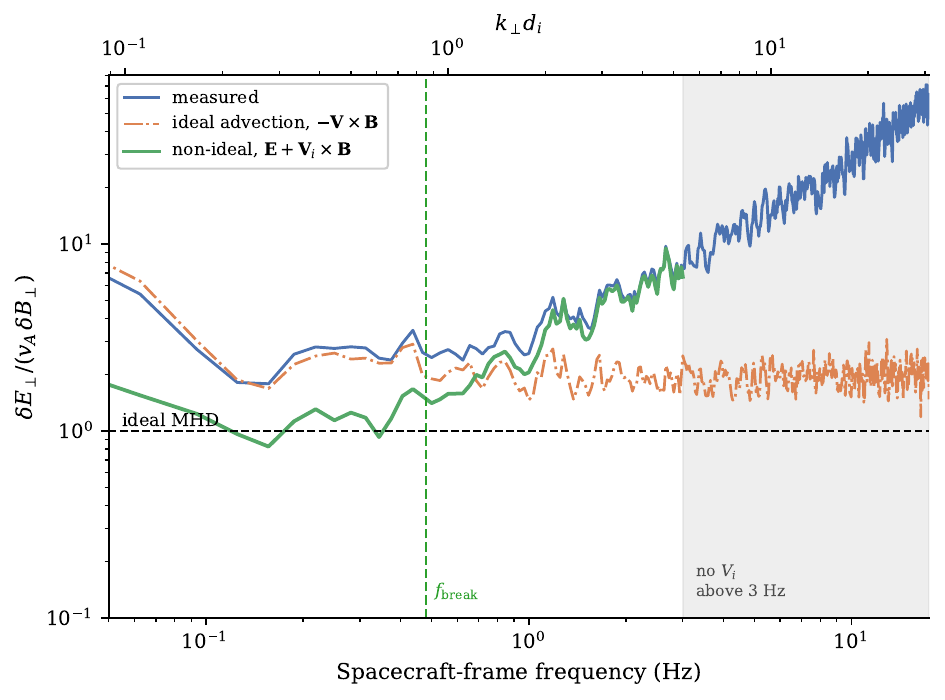}
\caption{Normalized electric-to-magnetic field ratio
$\dEperp/(\dBperp\vA)$ against spacecraft-frame frequency, with the
corresponding $\kperp d_i$ on the upper axis. The measured ratio (blue) is
separated into the ideal advection field $-\mathbf{V}\times\mathbf{B}$
(orange), which is flat, and the non-ideal remainder
$\mathbf{E}+\mathbf{V}_i\times\mathbf{B}$ (green), which is the part that
carries information about dispersion. Log--log slopes are $+0.92$ for the
measured ratio and $0.00$ for the advection field over 0.15--\vfgmScmHi~Hz, and
$+0.82$ for the remainder over 0.15--3~Hz, compared with the asymptotic KAW
value of $+1$ and zero for a non-dispersive mode. The dashed horizontal line marks the
ideal-MHD value of unity and the green dashed line the spectral break at
0.48~Hz. The remainder is not drawn above 3~Hz, where the 150~ms FPI cadence no
longer resolves $\mathbf{V}_i$. The EDP noise floor, measured from the
500--2000~Hz plateau of the instrument's own spectrum, is
$1.3\times10^{-4}$~mV$^{2}$~m$^{-2}$~Hz$^{-1}$, about a factor of 100 below the
signal at 10~Hz.}
\label{fig:ebspectrum}
\end{figure}

We therefore consider how the ratio varies with scale. The
dispersion relation of a wave mode can be recovered observationally from the
electric-to-magnetic field frequency spectrum \citep{chaston14}. A KAW
is distinguished from an ideal Alfv\'{e}n wave not by $E/B > 1$ alone but by an
$E/B$ that increases with $\kperp$. The measured ratio rises with a log--log
slope of $+0.92$, compared with the asymptotic KAW value of $+1$ and zero
for a non-dispersive mode. The advection field alone is flat to two decimal
places over the same range (Figure~\ref{fig:ebspectrum}). As an instrumental cross-check, ratios constructed
from the FGM and from the SCM magnetic spectra agree to within 5~\% at 2, 5 and
10~Hz, so the rise is not an artifact of the FGM response, and the
perpendicular electric-field signal remains approximately two orders of
magnitude above the measured EDP noise level at 10~Hz.

The measured ratio can also be compared with the two-fluid prediction, over
a limited range. For a KAW that relation gives
$\dEperp/(\vA\dBperp) = [1 + \kperp^2\rhoi^2(3/4 + T_e/T_i)]^{1/2}$
\citep{hollweg99,lysak96}. This is a finite-Larmor-radius expansion and holds
only for $\kperp\rhoi \lesssim 2$, a limit that at $\beta_i = 14.4$ is already
reached by 0.30~Hz, so we compare only below it. Over that range the measured
ratio is a median of \vtwoFluidRatio of the predicted value, approximately
\vtwoFluidOffset~\% below it. We do not treat this as a quantitative
confirmation of the relation. The expansion assumes cold ions and is being
applied at $\beta_i = 14$, it contains no Landau damping, and the kinetic-range
index of Section~\ref{sec:gauge} does not exclude weak damping of the cascade.
We therefore use this comparison as evidence for the expected scale dependence,
and not as a confirmation of the theoretical coefficient.

{}
We do not claim that the joint satisfaction of several criteria identifies
the mode on its own. None of these quantities is individually diagnostic, and
their conjunction is not a proof of uniqueness. The assessment set out in
Section~\ref{sec:discussion} rests instead on what each measurement can
constrain, together with a direct constraint on Hall-scale decoupling.

\begin{figure}[pos=htbp]
\centering
\includegraphics[width=\columnwidth]{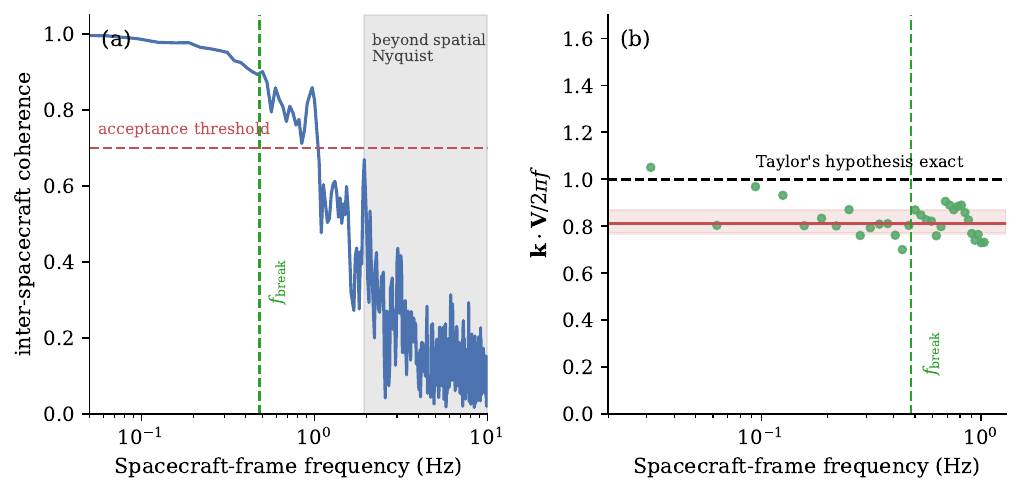}
\caption{Four-spacecraft test of Taylor's hypothesis by phase differencing
\citep{balikhin03}. (a)~Minimum inter-spacecraft coherence over the three
baselines. Frequencies are used only where it exceeds 0.7 and where the
wavenumber lies below the spatial Nyquist limit $k_{\max} = \pi/d_{\max}$
(shaded), which for the 33--41~km separations of this interval falls at
1.93~Hz. That leaves a usable band of 0.03--1.03~Hz, which contains the
spectral break (green dashed line). (b)~$\mathbf{k}\cdot\mathbf{V}/2\pi f$ over
that band, which is unity if the observed frequency is entirely a Doppler
shift. The solid line is the median, 0.81, and the shaded band its
interquartile range, \vkdotVloIQR--\vkdotVhiIQR. As a check of the method, a
synthetic plane wave of known $\mathbf{k}$, sampled at the measured spacecraft
positions, is recovered by the same analysis with a direction cosine of
1.000.}
\label{fig:multisc}
\end{figure}

Converting frequency to $\kperp$ throughout this paper uses the nominal
Doppler mapping $\kperp = 2\pi f/V_{\rm flow}$. We test this conversion
independently around the spectral break by phase differencing across the four
spacecraft \citep{balikhin03}, which gives a median
$\mathbf{k}\cdot\mathbf{V}/2\pi f = \vkdotV$ (interquartile range
\vkdotVloIQR--\vkdotVhiIQR), where exact agreement with Taylor's hypothesis
would give unity. The median differs from unity by 19~\%, indicating that the
Doppler conversion is supported at approximately the 20~\% level over the
resolved range around the break; we do not treat that departure as a formal
uncertainty. The method provides one phase-derived wave vector at each
frequency under the plane-wave assumption and does not determine the full
distribution of wave vectors in the turbulent fluctuations. The angle between
the bulk flow and $\mathbf{B}_0$ is $60^\circ$ (median; interquartile range
$36^\circ$--$77^\circ$), so the sampling direction is on average oblique to the
field, but that angle alone does not establish that the sampled wave vector is
perpendicular. Figure~\ref{fig:multisc} shows the coherence-based usable range
and the resulting ratio.

\section{Spectral Analysis and Dissipation Constraint}
\label{sec:spectra}

{}

\begin{figure}[pos=htbp]
\centering
\includegraphics[width=\columnwidth]{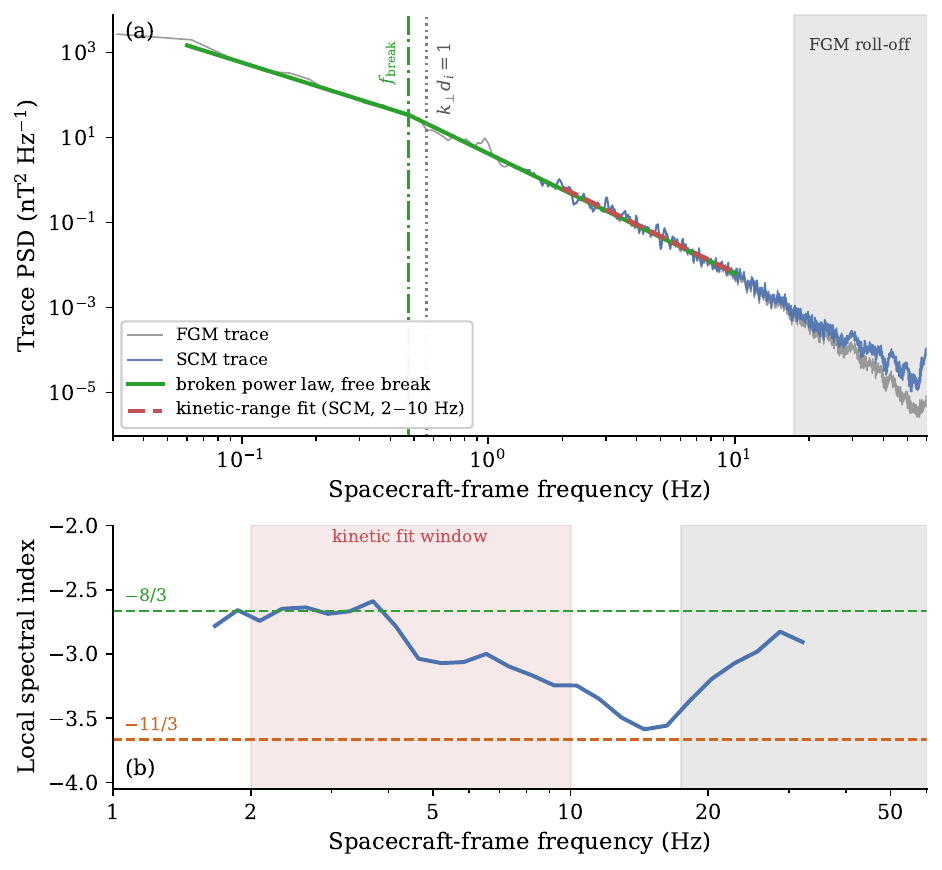}
\caption{{}
Trace magnetic power spectral density from MMS1 burst-mode FGM and SCM
data, Welch method with a Hanning window and 50~\% overlap, binned
logarithmically at 20 bins per decade.
(a)~The spectrum, with the broken power law fitted with the
break as a free parameter overplotted, giving
$\alpha_1 = \valphaFluid \pm \valphaFluidErr$ below the break and $f_{\rm break} = 0.48 \pm
0.06$~Hz; the kinetic-range index quoted in the text,
$\valphaKin \pm \valphaKinStat \pm \valphaKinWin$, is fitted to the SCM trace over 2--10~Hz. The
bright green vertical line marks $\kperp d_i = 1$ at $0.56$~Hz and the dashed
line the fitted break at $0.48$~Hz; the two agree within the fitted
uncertainty. The gray
band above 17~Hz marks the range over which FGM under-reports power relative
to SCM and is excluded from all fits.
(b)~The local spectral index of the SCM trace, fitted in a sliding window of
$\pm0.2$ decades, which is flat for a genuine power law. It steepens from about
$-2.7$ over 2--4~Hz to about $-3.5$ at 14~Hz, so no single index describes the
kinetic range; the shaded strip is the 2--10~Hz window over which the quoted
index is fitted, and the rise above 17~Hz lies inside the excluded band. Points
whose sliding window would extend beyond the fitted range are left undefined.}
\label{fig:psd}
\end{figure}

\subsection{Two-range spectral structure}
\label{sec:tworange}

Figure~\ref{fig:psd} shows the trace magnetic power spectral density (PSD)
computed from the burst-mode FGM and SCM data over the full 4.5-minute
interval. The spectrum shows a {}two-range structure separated by a
spectral break{}.
We locate the break by fitting a continuous broken power law with four
free parameters: the amplitude, the index below the break, the index above it,
and the break frequency. The break is therefore obtained without selecting a
separate narrow fitting window. The fit gives
{}
\begin{equation}
    \alpha_1 = \valphaFluid \pm \valphaFluidErr, \qquad
    \alpha_2 = -2.81 \pm 0.04, \qquad
    f_{\rm break} = \vfbreak \pm \vfbreakErr~\mathrm{Hz},
    \label{eq:brokenpl}
\end{equation}
and is stable over six combinations of Welch segment length (32 and 64~s) and
fitted range, with $\alpha_1$ between $-1.81$ and $-1.94$, $\alpha_2$ between
$-2.75$ and $-2.81$, and $f_{\rm break}$ between 0.459 and 0.506~Hz. The
fluid-scale index $\alpha_1 = \valphaFluid \pm \valphaFluidErr$
is noticeably steeper than the standard incompressible Kolmogorov
slope of $-5/3$, but is a well-documented characteristic of
magnetosheath turbulence.
The enhanced compressibility of post-shock plasma, the presence of
intermittent structures{}, and the more symmetric
distribution of Alfv\'{e}nic propagation directions compared to the
solar wind all act to modify the energy transfer rate and steepen the
{}spectrum below the break \citep{hadid17,hadid18,macek18}.
This steepening is a fluid-scale effect and is distinct from the
kinetic dissipation processes that operate above $f_{\rm break}$.

This range is not an inertial range in the conventional sense, and we do
not describe it as one. There is no decade-long power law below the break:
using FGM survey data over windows of up to 40~minutes, the band-averaged index
steepens from $-1.2$ over 0.02--0.06~Hz, to $-1.6$ over 0.06--0.2~Hz, to $-2.1$
over 0.2--0.5~Hz. These results are unchanged when the Welch segment length is
varied by a factor of 14 (64--900~s) and when linear detrending is switched
off, and the survey and burst products give the same indices over the same
266~s ($-1.50$ and $-1.91$ in the two bands), so the steepening is not an
artifact of the analysis. The range is also not stationary: consecutive
five-minute sub-intervals give indices of $-1.57$, $-0.66$, $-1.66$, $-1.14$
and $-1.95$ over 0.06--0.2~Hz. We therefore report $\alpha_1$ as a description
of the spectrum below the break and not as a scaling law, and no conclusion of
this paper depends on it.

Above the spectral break, in the kinetic range {}($2 < f <
10$~Hz), the
spectrum steepens further to {}$f^{-2.86}$.
The fitting window is set by three criteria: it lies above the ion-scale
break, below the electron-scale break at $\kperp\rhoe = 1$, which Taylor's
hypothesis places at 19.5~Hz, and inside the band over which FGM and SCM agree.
The third of these is the most restrictive. The two instruments agree to
within 15~\% between \vfgmScmLo and 12.4~Hz and to within 25~\% between
\vfgmScmLo and \vfgmScmHi~Hz; we retain a margin at each end and round inward,
which gives 2--10~Hz, and the upper limit is also a factor of two below the
electron-scale break. We find no evidence that FGM reaches its noise floor in
this interval: over 25--55~Hz the FGM trace does not exceed the SCM trace in
any of the 961 frequency bins, and the ratio of the two falls steadily to 0.12
at 60~Hz. Instead, FGM begins to under-report power above about 17~Hz as its
anti-aliasing response approaches the 64~Hz Nyquist frequency, which biases a
fitted index towards steeper values. The index quoted above is fitted to the
SCM trace; the FGM trace over the same window gives
$\valphaKinFGM \pm \valphaKinStat$.
We interpret this break as the transition from fluid to kinetic
dynamics{}, and
test whether it is more consistent with the ion inertial scale or with the ion
gyroscale.
{}In this
high-$\beta$ interval ($\beta_i = 14.4$) the two candidate scales are well
separated, $d_i = 45.0$~km against $\rhoi = 170.4$~km, and at
$\beta_i \gg 1$ the dispersive modification to Alfv\'{e}n waves
{}is expected at $k_\perp d_i \sim 1$ \citep{hasegawa76,leamon98}.
To evaluate this, we apply Taylor's frozen-in hypothesis
\citep{taylor38}, which relates the spacecraft-frame frequency $f$
to the plasma-frame perpendicular wavenumber via
$\kperp = 2\pi f / V_{\rm flow}$.
Because the break is now fitted rather than assumed, we state the scale it
corresponds to directly. With $f_{\rm break} = \vfbreak \pm \vfbreakErr$~Hz and
$V_{\rm flow} = \vVflow \pm \vVflowErr$~km~s$^{-1}$,
\begin{equation}
    \kperp d_i = \vkdiBreak \pm \vkdiBreakErr, \qquad
    \kperp \rhoi = \vkrhoiBreak \pm \vkrhoiBreakErr ,
    \label{eq:kdi}
\end{equation}
The flow speed implied by $\kperp d_i = 1$ at the fitted break is
$2\pi f_{\rm break} d_i \approx 136 \pm 18$~km~s$^{-1}$, consistent with the measured
$\vVflow \pm \vVflowErr$~km~s$^{-1}$.
{}
A useful property of the Taylor mapping is that spectral indices
are preserved: if $P(k) \propto k^{\alpha}$ in the plasma frame,
then $P(f) \propto f^{\alpha}$ in the spacecraft frame, so the
observed indices {}$\alpha_1$ and $\alpha_2$ translate directly into
wavenumber-space slopes for comparison with theory.
We note that with $V_{\rm flow} / \vA = 2.1$, the Taylor frozen-in
condition is only marginally satisfied. However, at sub-ion scales
($\kperp d_i > 1$), the perpendicular phase speed of KAWs remains
small compared with the bulk flow, so perpendicular sweeping
continues to dominate the frequency-to-wavenumber mapping
\citep{howes14b,klein14b}. In this moderate-flow regime, the
partial breakdown of Taylor's hypothesis primarily introduces minor
frequency broadening rather than a fundamental shift in the location
of the spectral break.
The four-spacecraft measurement in Section~\ref{sec:identification} bears
directly on this. Phase differencing gives
$\mathbf{k}\cdot\mathbf{V}/2\pi f = \vkdotV$, so the departure from exact
frozen-in advection is about 20~\%. The uncertainties quoted for $\kperp d_i$
and $\kperp\rhoi$ in Equation~(\ref{eq:kdi}) reflect the fitted-break
uncertainty alone, and the four-spacecraft result indicates an additional
limitation of approximately 20~\% in the Doppler conversion. Even with that
limitation, the measured break remains substantially closer to
$\kperp d_i = 1$ than to $\kperp\rhoi = 1$, the two scales differing here by a
factor of \vrhoiOverdi.
{}
We therefore describe the fitted break at $\vfbreak \pm \vfbreakErr$~Hz as
more consistent with the ion inertial scale than with the ion gyroscale for
this interval.

\subsection{{}Kinetic-range
index as a constraint on the cascade}
\label{sec:gauge}

The kinetic-range spectral index of {}$-2.86 \pm
0.06\,(\mathrm{stat}) \pm 0.18\,(\mathrm{fit\ window})$ can be placed in context
using two theoretical limiting cases established in the literature.
The second uncertainty is the spread obtained over plausible fitting
windows, defined as those at least half a decade wide, beginning above the ion
break and ending below the electron break. It exceeds the statistical error and
is therefore quoted alongside it; combined in quadrature the two give
$\pm\valphaKinComb$. The kinetic range is curved rather than a single power
law: the local index steepens from about $-2.7$ at 2--4~Hz to about $-3.5$ at
14~Hz, so any single index is a summary of that profile and not a scaling law
(Figure~\ref{fig:psd}b).
\begin{figure}[pos=htbp]
\centering
\includegraphics[width=0.93\columnwidth]{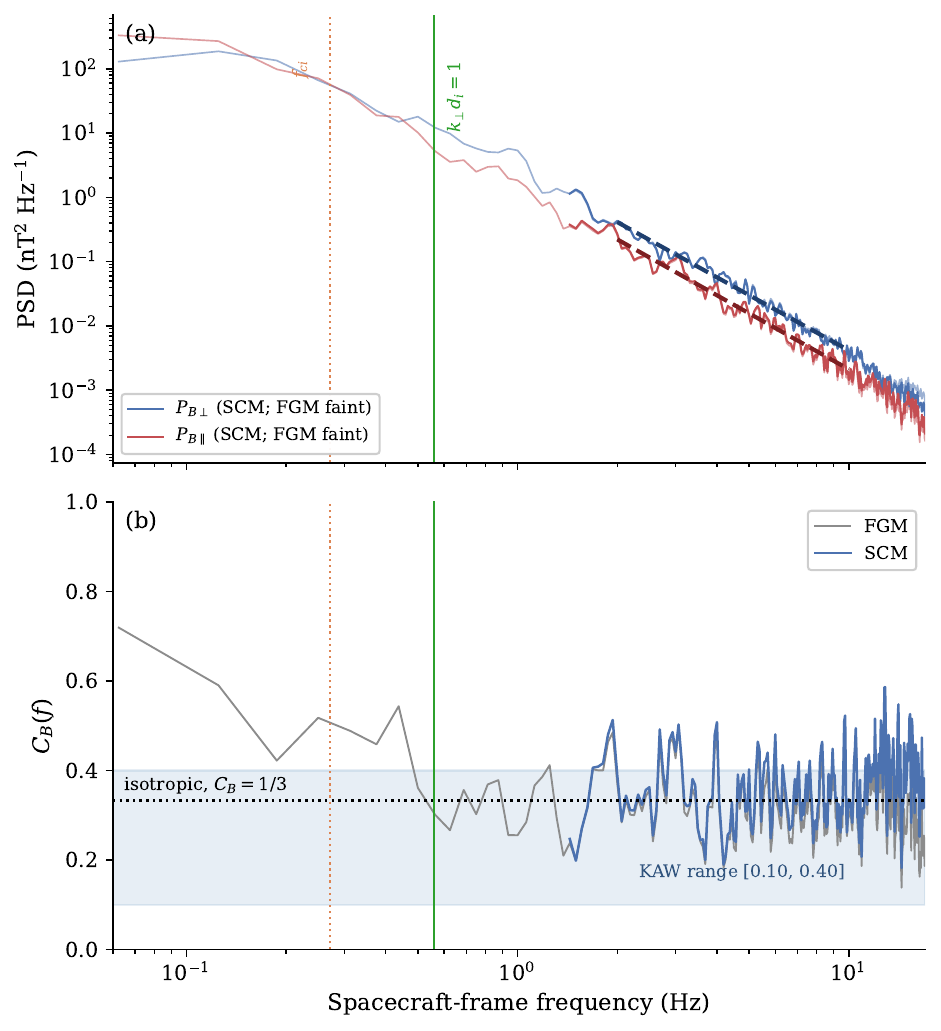}
\caption{{}Spectral anisotropy ratio and compressibility from MMS1 burst-mode
{}data decomposed in field-aligned coordinates (FAC).
(a)~Perpendicular ($P_{B\perp}$, blue) and parallel ($P_{B\parallel}$,
red) magnetic power spectra, with kinetic-range power-law fits
($\alpha^{\rm kin}_{\perp} = {}\valphaPerp \pm \valphaPerpErr$ in firebrick,
$\alpha^{\rm kin}_{\parallel} = {}\valphaPar \pm \valphaParErr$ in navy)
overplotted above the ion-scale break, fitted from SCM over 2--10~Hz with
logarithmic binning.
The {}fluid-scale index is reported in Figure~\ref{fig:psd}.
(b)~Magnetic compressibility $C_B(f) = P_{B\parallel}/(P_{B\parallel}
+ P_{B\perp})$; the blue-shaded band marks the theoretical KAW range
$C_B \in [0.10, 0.40]$ \citep{sahraoui09,stasiewicz00}, with the
upper bound explicitly marked by a horizontal dashed line for
visual reference.
The horizontal dotted line marks the isotropic value $C_B = 1/3$.
In both panels the green vertical line marks $\kperp d_i = 1$
(0.56~Hz) and the orange dotted line marks $f_{ci} = 0.27$~Hz.
The display is truncated at 17~Hz, the upper limit of the verified FGM
range.
Above the ion-scale break, $P_{B\perp}/P_{B\parallel} \approx {}2.22$
{}and $C_B \approx 0.31$ (median),
{}
both close to the isotropic values of 2.00 and $1/3$. This does not
support a mirror-mode-dominated interval ($C_B \sim 0.8$--$1.0$), but it does
not constrain whistler modes, which are addressed by the wave-normal angle and
circular-polarization measure in Section~\ref{sec:wavenormal}.}
\label{fig:polarization}
\end{figure}
\noindent In the absence of significant collisionless dissipation, the KAW
turbulent cascade is governed by dispersive nonlinear interactions.
Theoretical work \citep{boldyrev12,howes08} and observations \citep{sahraoui09,sahraoui10}
consistently find a kinetic-range spectrum scaling as
$k_{\perp}^{-8/3} \approx k_{\perp}^{-2.67}$ in this conservative
regime. At the other extreme, when Landau damping strongly suppresses the
cascade, the spectrum steepens toward $k_{\perp}^{-11/3} \approx
k_{\perp}^{-3.67}$ \citep{passot15,alexandrova12}.

The observed index {}$\valphaKin \pm \valphaKinStat\,(\mathrm{stat}) \pm
\valphaKinWin\,(\mathrm{fit\ window})$ lies much closer to the weakly damped
$-8/3$ limiting prediction than to the strongly damped $-11/3$ limiting
prediction, and therefore disfavors the strongly damped limiting description
for this interval. The second uncertainty is a spread over plausible fitting
windows rather than a statistical standard deviation, so we do not combine the
two into a single significance; and because the kinetic range is curved we use
the index only to constrain the cascade.
{}Within these limiting descriptions, $-8/3$ represents
weaker attenuation of the cascade and $-11/3$ stronger damping.
{}
{}
We caution that the spectral index alone cannot distinguish between
Landau damping and other collisionless dissipation channels (e.g.\
stochastic heating, transit-time damping); it {}places the cascade relative to the two
theoretical limits, and determines neither the amount of particle heating nor
the responsible mechanism.
The large $\beta_i = 14.4$ is notable in this context: at
$\beta_i \gg 1$ the resonant condition $v_{\parallel} \approx
\omega/k_{\parallel} \approx \vA$ lies well within the ion thermal
distribution, making wave-particle interactions kinematically
accessible to a large fraction of the particle population
\citep{howes14,tenbarge13}.

{}
Landau
damping, transit-time damping and stochastic heating all steepen a cascade, and
separating them requires velocity-space measurements such as the
field--particle correlation analysis of \citet{afshari21}, which is beyond what
electromagnetic data can provide.

The observed spectral properties of this interval are summarized in
Table~\ref{tab:spectral}.

\begin{table}[width=.9\linewidth,pos=ht]
\caption{Observed magnetic spectral properties, MMS1,
2015 December 28. The implied-flow uncertainty uses the unrounded
break-frequency uncertainty of 0.062~Hz, with $d_i$ held fixed.}
\label{tab:spectral}
\begin{tabular*}{\tblwidth}{@{} LC @{}}
\toprule
Property & Observed value \\
\midrule
{}
Fluid-scale index $\alpha_1$ & $\valphaFluid \pm \valphaFluidErr$ \\
Kinetic-range index             & {}$\valphaKin \pm \valphaKinStat \pm \valphaKinWin$ \\
Spectral break frequency        & {}$\vfbreak \pm \vfbreakErr$~Hz \\
Corresponding plasma scale      & {}$\kperp d_i = \vkdiBreak \pm \vkdiBreakErr$ \\
Ion gyroscale at the break & $\kperp \rhoi = 3.23 \pm 0.42$ \\
Implied flow velocity & {}$136 \pm 18$~km~s$^{-1}$ \\
Measured bulk flow              & $159 \pm 33$~km~s$^{-1}$ \\
Taylor test, $\mathbf{k}\cdot\mathbf{V}/2\pi f$ & $0.81$ \\
Wave-normal angle $\theta_{kB}$ & $75^\circ$ \\
Circular-polarization measure $\sigma_B$ & $+0.02$ \\
Plasma $\beta_i$                & $14.4$ \\
\bottomrule
\end{tabular*}
\end{table}

\subsection{{}Spectral anisotropy ratio and compressibility}
\label{sec:polarization}

Figure~\ref{fig:polarization} provides an independent {}characterization of the magnetic spectral anisotropy and compressibility
by decomposing the burst-mode FGM magnetic fluctuations in field-aligned
coordinates (FAC) and computing the magnetic compressibility
\begin{equation}
    C_B(f) = \frac{P_{B\parallel}}{P_{B\parallel} + P_{B\perp}},
    \label{eq:cb}
\end{equation}
where $P_{B\parallel}$ and $P_{B\perp}$ are the power spectral densities
of field-parallel and field-perpendicular magnetic fluctuations,
respectively.
The FAC basis is constructed from a $10\,\mathrm{s}$ moving average of the measured
field, identical to the background subtraction used in
Section~\ref{sec:data}.

Panel~(a) of Figure~\ref{fig:polarization} shows $P_{B\perp}(f)$ and
$P_{B\parallel}(f)$ separately.
{}Below the break the two components carry approximately equal
power ($P_{B\perp}/P_{B\parallel} \approx 1.0$), consistent with
isotropic large-scale MHD Alfv\'{e}nic turbulence.
Above the ion-scale break at $\kperp d_i = 1$ (0.56~Hz, green line),
{}the median ratio is
$P_{B\perp}/P_{B\parallel} = \vPratio$ from FGM and $\vPratioscm$ from SCM over
1.4--10~Hz.
Both values require an isotropic baseline for comparison. Three
uncorrelated components of equal power give
$P_{B\perp}/P_{B\parallel} = 2.00$ and $C_B = 1/3$ exactly, so a ratio of two
is the value produced by isotropy. The measured values are therefore not
evidence of transverse dominance, and the kinetic-range fluctuations are better
described as close to the corresponding isotropic values.
Panel~(b) shows the magnetic compressibility $C_B(f)$.
The median value in the kinetic range is $C_B = {}\vCB$
from FGM and $\vCBscm$ from SCM, which falls
within the theoretical KAW prediction of $C_B \in [0.10, 0.40]$
\citep{sahraoui09,stasiewicz00}.
This quantity and the ratio above are both defined against the
background-field direction, and they move by 32~\% and 53~\% respectively when the
averaging window is varied by a factor of 20. Neither excursion changes what
they constrain; the full sensitivity is reported in
Section~\ref{sec:discussion}.
{}
Since that band straddles the isotropic value, $C_B$ is a consistency
check here and not a discriminant. It does, however, argue against a
mirror-mode-dominated interval, since mirror modes are purely compressive with
$C_B \sim 0.8$--$1.0$. It does not constrain whistler-mode waves: a
quasi-parallel whistler is transverse, so $C_B \to 0$, which lies inside the
same band. The wave-normal angle and circular-polarization measure measured
below provide
the relevant constraint instead.
The transition from moderately compressive ($C_B \approx 0.4$--$0.7$,
{}fluid-scale range) to low-compressibility ($C_B \approx
0.31$, kinetic
range) occurs {}around the spectral break (Figure~\ref{fig:psd}).
The spectral shape is stable even though the fluctuation amplitude is
not. Recomputing both quantities in each 66~s quarter gives
kinetic indices of $-2.81$, $-2.87$, $-2.86$ and $-2.97$. That spread, 0.16,
is smaller than the $\pm\valphaKinComb$ combined uncertainty quoted in
Section~\ref{sec:gauge}.
$C_B$ stays between 0.22 and 0.37, inside the KAW
range in every quarter, while the rms fluctuation amplitude increases from
\vrmsBqLo to \vrmsBqHi~nT across the same quarters. A whole-interval power
spectrum is therefore reasonable for the kinetic-range analysis presented here.
This applies only to the kinetic range and does not imply that the fluid-scale
spectrum is stationary.

\subsection{Wave-normal angle and circular-polarization measure}
\label{sec:wavenormal}

The quantities above constrain how magnetic energy is partitioned between
directions, but not the direction of propagation. A large wave-normal angle is
one of the defining signatures of the KAW mode, and the angle at which the
transition to KAW behavior occurs depends on plasma $\beta$ and on composition
\citep{villarroel23}. We have therefore computed the wave-normal angle
directly. The magnetic spectral matrix is formed from a
Morlet wavelet transform of the three components and averaged over eight wave
periods, and the wave normal is obtained from the singular value decomposition
of the $6\times3$ real matrix built from its real and imaginary parts
\citep{santolik03}. We report
$\theta_{kB} = \arccos|\hat{\mathbf{k}}\cdot\hat{\mathbf{b}}_0|$ on
$[0,90]^\circ$; a single spacecraft does not determine the sign of
$\mathbf{k}$, so that $\theta_{kB}$ is unaffected by the Doppler shift. Cells
are retained only where the normal is well determined, using planarity and
degree of polarization both above 0.7, and \vqualityPass~\% of cells pass. The
background field is the local mean over ten periods of the scale being
analyzed, as is standard for anisotropy measurements \citep{horbury08,chen11}.

\begin{figure}[pos=htbp]
\centering
\includegraphics[width=\columnwidth]{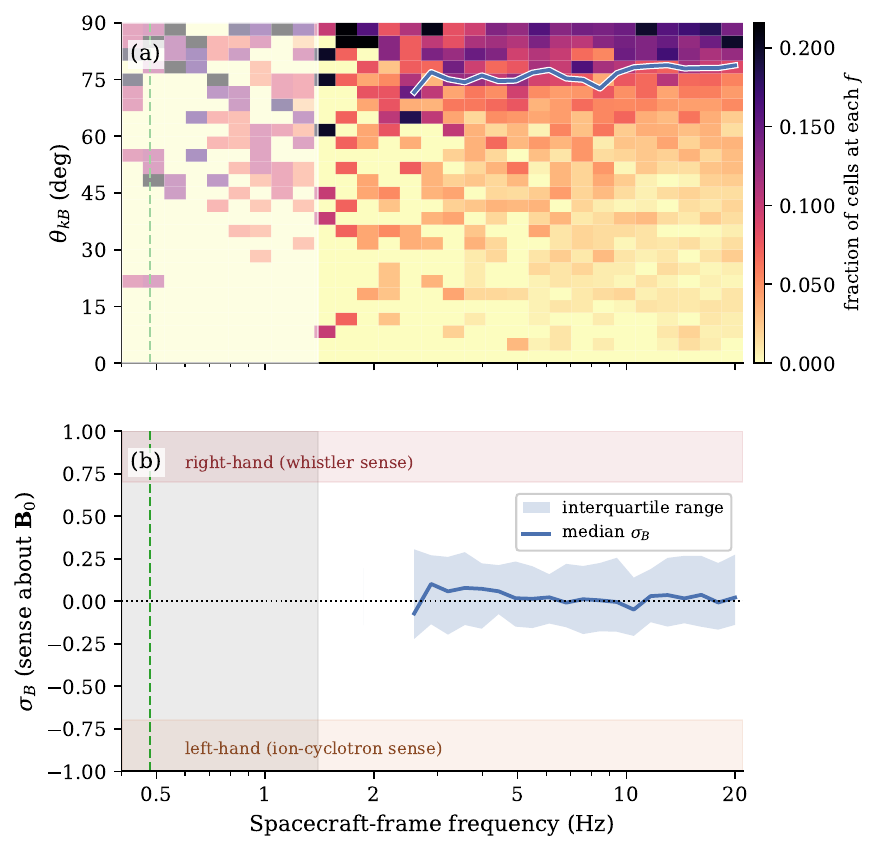}
\caption{Wave-normal angle and circular-polarization measure for MMS1, from the
singular value decomposition of the wavelet spectral matrix
\citep{santolik03}. (a)~Distribution of $\theta_{kB}$ against spacecraft-frame
frequency, normalized at each frequency so that every frequency contributes
equally; the blue line is the median. (b)~Median transverse circular
polarization $\sigma_B$ about $\mathbf{B}_0$ with its interquartile range,
positive for right-hand rotation. The shaded bands mark $|\sigma_B| > 0.7$, the
whistler and ion-cyclotron senses. In both panels the green dashed line marks
the spectral break at 0.48~Hz, and the pale band below 1.4~Hz marks
frequencies excluded from the quoted statistics because FGM and SCM agree only
above it. Over 1.4--10~Hz the median angle is $75^\circ$ (interquartile range
$59^\circ$--$84^\circ$) and $\sigma_B = +0.02$ (interquartile range $-0.17$ to
$+0.23$). These band values pool all accepted cells over that range; the median
curves are drawn only where at least 25 cells are accepted. The wave normal is
well determined in \vqualityPass~\% of cells.}
\label{fig:wavenormal}
\end{figure}

Over 1.4--10~Hz the wave-normal angle is $\theta_{kB} = 75^\circ$
(interquartile range $59^\circ$--$84^\circ$) from SCM and $76^\circ$ from FGM.
It varies little across the spectrum, taking values of $70^\circ$, $67^\circ$,
$73^\circ$ and $72^\circ$ in the bands 0.4--0.8, 0.8--1.4, 1.4--3 and
3--10~Hz, so the broad-band medians show no abrupt change in propagation angle
around the spectral break
(Figure~\ref{fig:wavenormal}a).
The measured value is not $90^\circ$, and is not expected to be. We also
passed synthetic magnetic
fluctuations of known geometry through the same analysis, which provides two
reference points. An ensemble with
$\mathbf{k}$ strictly perpendicular to $\mathbf{B}_0$ is recovered at
$78^\circ$, and randomly oriented wave normals give $59^\circ$, not $45^\circ$,
because random directions follow a $\sin\theta$ distribution. The measured
$75^\circ$ lies close to the first and well above the second, and 36~\% of cells
exceed $80^\circ$, compared with 17~\% for randomly oriented normals. The measurement
therefore supports quasi-perpendicular propagation, as expected for KAWs.

The circular-polarization measure provides the constraint on the whistler
interpretation that $C_B$ cannot. The transverse circular polarization about
$\mathbf{B}_0$ is
$\sigma_B = +0.02$ (interquartile range $-0.17$ to $+0.23$;
Figure~\ref{fig:wavenormal}b), that is centred close to zero: 70~\% of cells
have $|\sigma_B| < 0.3$ and only 1.7~\% reach the $\sigma_B > +0.7$ expected of
a whistler. The large wave-normal angle does not support quasi-parallel
propagation, while the polarization result does not support a strongly
right-hand-polarized whistler interpretation. We therefore use the combined
angle and polarization measurements as evidence for quasi-perpendicular
KAW-like fluctuations, and not as a unique mode identification. Two limitations
apply to this measurement. The wave normal is well determined in only
\vqualityPass~\% of cells, and the method saturates below $90^\circ$, so we do
not infer a precise $k_\parallel/\kperp$ ratio and limit our conclusion to
quasi-perpendicular propagation.

\section{Discussion and Conclusions}
\label{sec:discussion}

\paragraph{{}Consistency with KAW-like turbulence:}
{}
Taken together, the measurements in this interval are consistent with
quasi-perpendicular KAW-like turbulence. The electric-to-magnetic field ratio
rises with wavenumber, as expected for a dispersive mode and not for a
structure carried past the spacecraft; the wave-normal angle is large and the
circular-polarization measure is centred near zero; the compressibility and the
ion temperature anisotropy do not support a mirror-mode-dominated interval; and
the ion and electron frozen-in residuals separate near $\kperp d_i \sim 1$.
Each of these bears on a different candidate interpretation, as collected in
Table~\ref{tab:exclusions}, and none is uniquely diagnostic on its own. We
therefore describe the fluctuations as KAW-like rather than claiming a unique
mode identification.

\begin{table}[width=\linewidth,pos=ht]
\caption{Measurements used to constrain candidate interpretations in this
interval.}
\label{tab:exclusions}
\begin{tabular*}{\tblwidth}{@{} LLL @{}}
\toprule
Observable & Measured & Interpretive constraint \\
\midrule
Wave-normal angle $\theta_{kB}$ & $75^\circ$ & supports quasi-perpendicular propagation \\
Circular-polarization measure $\sigma_B$ & $+0.02$ & no strong right-hand polarization \\
Magnetic compressibility $C_B$ & $0.31$ & does not support mirror-mode dominance \\
$T_\perp/T_\parallel$; $r(|B|,n_i)$ & $0.98$; $-0.30$ & median below mirror threshold; weak pressure-balanced component \\
$|B|$ enhancements & $< 2.36\times$ background & no sustained SLAMS-like or shocklet-like enhancements \\
$|E{+}V_e{\times}B|/|E{+}V_i{\times}B|$ & $\vfrozenRatio$ & consistent with Hall-scale decoupling; not purely ideal advection \\
\bottomrule
\end{tabular*}
\end{table}

\begin{figure}[pos=htbp]
\centering
\includegraphics[width=\columnwidth]{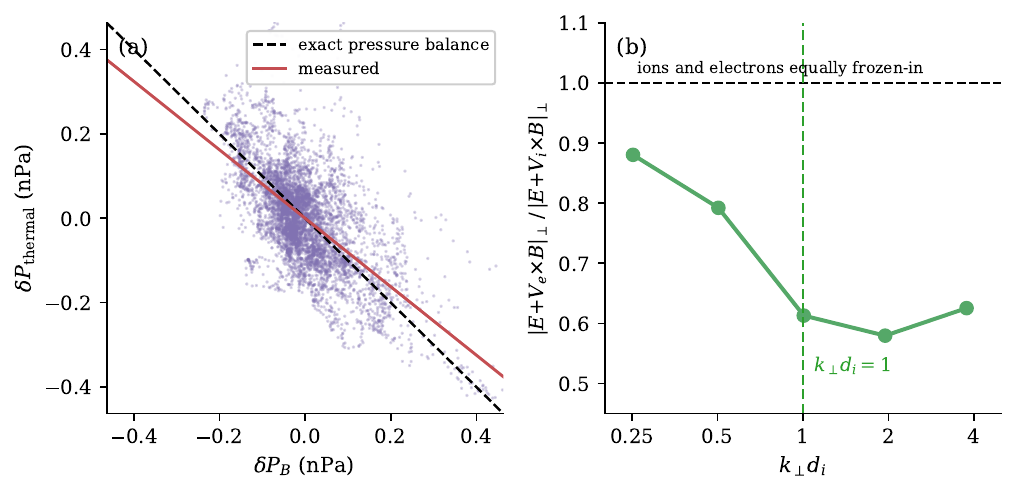}
\caption{(a)~Thermal against magnetic pressure fluctuations about a 10~s
running mean. Exact pressure balance would follow the dashed line of slope
$-1$; the measured regression slope is $\vpressureSlope$ with $r = \vpressureR$, so a
pressure-balanced component is present but does not account for the
fluctuations. (b)~Ratio of the perpendicular electron to ion frozen-in
residuals, $|\mathbf{E}+\mathbf{V}_e\times\mathbf{B}|_\perp /
|\mathbf{E}+\mathbf{V}_i\times\mathbf{B}|_\perp$, in five wavenumber bands.
Both residuals vanish for a structure advected with the plasma, and the ratio
is unity when ions and electrons are equally frozen to the field. It falls from
0.88 at $\kperp d_i = 0.25$ to 0.61 at $\kperp d_i = 1.0$ and stays near 0.6
above, which is the expected Hall signature at the ion inertial scale. Both
panels use FPI moments at 150~ms cadence for the ions and 30~ms for the
electrons.}
\label{fig:frozenin}
\end{figure}

The final row provides a direct constraint on Hall-scale ion--electron
decoupling. At $\kperp d_i \sim 1$ the magnetic
field decouples from the ions but not from the electrons, so
$\mathbf{E} + \mathbf{V}_i\times\mathbf{B}$ is the Hall and
electron-pressure term of the generalized Ohm's law \citep{vasyliunas75}, which
is what makes a KAW dispersive, while
$\mathbf{E} + \mathbf{V}_e\times\mathbf{B}$ remains small
\citep{mozer02}. Both vanish for a
structure advected with the plasma. Using FPI moments, the rms of the
perpendicular electron residual is 0.455~mV~m$^{-1}$ compared with
0.609~mV~m$^{-1}$ for the ion residual, a ratio of 0.74. Resolved by scale the
ratio is 0.88 at $\kperp d_i = 0.25$, falls to 0.61 at $\kperp d_i = 1.0$, and
remains near 0.6 above it, so the field remains more closely frozen to the
electrons than to the ions near the ion inertial scale. This is the expected
signature of Hall-scale physics, it occurs at $d_i$, and it does not support an
interpretation in terms of a purely ideal structure advected with the plasma.
Instrument noise reduces correlations and therefore drives this ratio
towards unity, so the measurement is a conservative one. Because this test uses
FPI bulk moments it is not based on field data alone; it does not, however,
require particle distribution functions or field--particle correlation
analysis. Figure~\ref{fig:frozenin}(b) shows the ratio against $\kperp d_i$.

We note also that we do not treat Hall-MHD turbulence, sometimes listed as
an alternative to KAW turbulence, as a separately excluded wave mode. At
$\kperp d_i \gtrsim 1$ the kinetic Alfv\'{e}n wave is the normal mode of
Hall-MHD, and the measured ion--electron decoupling is the Hall-scale property
relevant to the KAW interpretation. The alternative that the term does point to is a
non-propagating Hall structure such as an advected current sheet, which the
frozen-in test above does not support.

\paragraph{{}Cascade state and spectral index:}
The kinetic-range index {}$\valphaKin \pm \valphaKinStat \pm \valphaKinWin$ provides an observational constraint
on the {}state of the cascade.
{}As set out in Section~\ref{sec:gauge}, it lies much
closer to the weakly damped $-8/3$ limiting prediction than to the strongly
damped $-11/3$ one, and therefore disfavors the strongly damped limiting
description for this interval. Because the kinetic range is curved and several
collisionless processes can steepen a spectrum, the spectral index
constrains the cascade, not the dissipation mechanism.
The high $\beta_i = 14.4$ is relevant in this context: at $\beta_i \gg 1$
the ion thermal velocity substantially exceeds $\vA$, so
wave-particle resonance conditions are kinematically accessible to a
large fraction of the particle population \citep{howes14,chen19,horvath20}.
Whether this results in preferential ion or electron energization
cannot be determined from electromagnetic data alone{};
\citet{afshari21} report field--particle correlation evidence of electron
Landau damping in magnetosheath intervals, which is the class of measurement
required to settle it.
We {}emphasize that a spectral index alone does not identify the specific
dissipation channel, and cannot; Landau damping, stochastic heating, and
transit-time damping all produce cascade steepening and cannot be
distinguished from this measurement.
Landau damping in particular is a velocity-space process and cannot be
inferred from a spectral slope; confirming it requires distribution function
measurements.
A useful extension of this approach would be to compile kinetic-range
spectral indices from many MMS magnetosheath intervals and correlate
them with local $\beta_i$ and independently measured dissipation
proxies such as the field-particle correlation signatures of
\citet{afshari21}, building an empirical calibration between the
spectral index and the dissipation level across varying plasma
conditions.
{}
Any such survey would need to quote the fit-window systematic alongside
the statistical error, since the two are comparable here.

\paragraph{{}Fluid-scale steepening:}
The {}fluid-scale slope {}$\valphaFluid \pm \valphaFluidErr$, steeper than Kolmogorov $-5/3$,
should not be attributed to kinetic dissipation.
The steepening {}below the break is a
well-understood fluid-scale consequence of compressibility,
intermittency, and bow-shock imprinting on the downstream plasma
\citep{hadid17,hadid18,macek18}.
The spectral break at {}$\vfbreak \pm \vfbreakErr$~Hz {}separates the
{}fluid-scale and kinetic {}regimes.
The two ranges respond to different physics: compressibility
modifies the {}fluid-scale range, while {}dispersive kinetic dynamics govern the kinetic range.
This separation is evident in the spectral structure and should be
borne in mind when comparing magnetosheath spectral indices with
solar wind or numerical simulation results.

\paragraph{High-$\beta$ context and kinetic scale identification:}
The high ion beta $\beta_i = 14.4$ of this interval places it at the
upper end of magnetosheath conditions \citep{lucek05,macek18} and has
direct consequences for the scale identification.
When $\beta_i \gg 1$, the ion inertial length $d_i$ and the ion
gyroradius $\rhoi$ are related by $\rhoi = \sqrt{\beta_i}\,d_i$;
here $\rhoi = 3.79\,d_i = 170.4$~km versus $d_i = 45.0$~km.
The {}fitted
break gives $\kperp d_i = \vkdiBreak \pm \vkdiBreakErr$ and
$\kperp\rhoi = \vkrhoiBreak \pm \vkrhoiBreakErr$. For this interval the measured
break is therefore more consistent with the ion inertial scale than with the
ion gyroscale,
consistent with the theoretical expectation that at $\beta \gg 1$,
dispersive effects in KAW physics first appear at $k_\perp d_i \sim 1$
\citep{hasegawa76,stasiewicz00}.
This distinction matters for future surveys: at typical magnetosheath
$\beta \sim 1$--$5$, $d_i$ and $\rhoi$ differ by factors of 1--2
and may both contribute to spectral features; at $\beta \sim 14$,
{}the two are separated well enough to
test which of them the break follows, which is valuable for
constraining the underlying wave physics.

\paragraph{Limitations:}
This {}study
examines one 4.5-minute interval. Most diagnostics use MMS1
measurements, while the phase-differencing test uses all four spacecraft.
While the interval is representative of dayside magnetosheath
conditions and has been independently validated in prior studies
\citep{macek18,stawarz23}, a statistical survey of many MMS
magnetosheath intervals at different $\beta$ values, solar wind
conditions, and magnetosheath locations (subsolar, flank) would
substantially strengthen the conclusions and allow the spectral
{}index to be calibrated against local plasma parameters.
Such an extension is a natural and straightforward application of the
methodology developed here.

Four limitations remain. First, a subdominant
pressure-balanced component is present: comparing the magnetic and thermal
pressure fluctuations gives $r = \vpressureR$ with a regression slope of $-0.81$,
compared with the $r = -1$ and slope $-1$ of exact pressure balance
(Figure~\ref{fig:frozenin}a). This is
consistent with the weak $|B|$--$n_i$ anticorrelation reported in
Section~\ref{sec:data}. At $\beta_i \approx 14$ almost any low-frequency mode
tends towards approximate pressure balance, KAWs included, so this does not
argue against the identification. We therefore acknowledge that slow-mode-like
or pressure-balanced fluctuations may contribute to the interval; pressure
balance alone does not distinguish them from KAW fluctuations in this
high-$\beta$ plasma.

Second, four-spacecraft phase differencing provides one phase-derived wave
vector at each frequency, under the assumption of a single plane wave or of a
dominant coherent component, with the three independent baselines supplying the
three components of that vector. The method does not recover the full
distribution of simultaneous wave vectors present in broadband turbulence. We
therefore use $\mathbf{k}\cdot\mathbf{V}/2\pi f$ only to test the Doppler
conversion around the spectral break, while the distribution of wave-normal
angles is assessed separately by the single-spacecraft analysis of
Section~\ref{sec:wavenormal}. Recovering the full distribution of $\mathbf{k}$
would require $k$-filtering \citep{pincon91}, which needs more independent
realizations than a 266~s burst provides.

Third, we tested the sensitivity of every background-dependent quantity to
the averaging window. For fixed windows from 3 to 60~s, $C_B$ varies from 0.254
to 0.357, $P_{B\perp}/P_{B\parallel}$ from 1.80 to 2.94, the median $E/B$ ratio
from 2.17 to 3.90, and the rms $\dEpar$ from 0.60 to 0.70~mV~m$^{-1}$; for the
wave-normal calculation, fixed windows between 3 and 30~s give median angles
between $\vthetaWinLo^\circ$ and $\vthetaWinHi^\circ$. Over the same windows
the measured $E/B$ slope stays between $+0.91$ and $+0.94$ and the
advection-field slope between $-0.01$ and $+0.01$. The numerical values
therefore depend on the background definition, but no conclusion reverses:
$C_B$ stays inside $[0.10, 0.40]$ for every window and far below the
mirror-mode range, $\theta_{kB}$ stays above the $59.3^\circ$ obtained from
randomly oriented normals, and the scale-dependent $E/B$ behaviour, the finite
parallel electric field, and the quasi-perpendicular propagation are unchanged.
The spectral indices and the break frequency are obtained from the trace
spectrum and do not depend on the background-field direction at all.

Fourth, the fluid-scale range is neither a power law nor stationary, so
$\alpha_1$ is reported as a description of the spectrum below the break and no
conclusion rests on it.

{}
\label{sec:conclusion}

\paragraph{Conclusions:}
Using {}MMS burst-mode electromagnetic measurements in the dayside
magnetosheath on 2015 December~28, we {}conclude the following:

\begin{enumerate}
    \item {}
    The combined measurements are consistent with quasi-perpendicular
    KAW-like turbulence, but they do not provide a unique
    mode identification. The wave normal is quasi-perpendicular, with
    $\theta_{kB} = 75^\circ$ compared with $78^\circ$ for a strictly
    perpendicular ensemble analyzed in the same way and $59^\circ$ for randomly
    oriented normals, and the circular-polarization measure is centred near
    zero ($\sigma_B = +0.02$), which does not support a strongly
    right-hand-polarized whistler interpretation. The compressibility, the ion
    anisotropy and the pressure-balance results do not support a
    mirror-mode-dominated interval, although a weak pressure-balanced
    contribution is present, and no sustained SLAMS-like or shocklet-like
    magnetic enhancements are found. The ion and electron frozen-in residuals
    differ by 26~\%, consistent with Hall-scale decoupling and not with a purely
    ideal structure advected with the plasma. We do not claim that these
    criteria are individually necessary or jointly sufficient for a KAW.

    \item The fluctuations are dispersive. The electric-to-magnetic field
    ratio rises with wavenumber, with a log--log slope of $+0.92$, compared
    with the KAW asymptotic value of $+1$ and zero for any non-dispersive mode. The
    advection field $-\mathbf{V}\times\mathbf{B}$ is flat to two decimal
    places over the same range, so an elevated $E/B$ ratio on its own is mostly
    a consequence of $V_{\rm flow}/\vA = 2.1$ rather than evidence of
    dispersion. We therefore use the scale dependence of the ratio, rather than
    its value at a single scale, as evidence of dispersive behaviour. With the
    advection field removed the ratio is \vEBnonideal.

    \item {}
    For this interval the measured spectral break is more consistent with
    the ion inertial scale than with the ion gyroscale. Fitting a broken power
    law with the break as a free parameter gives
    $f_{\rm break} = \vfbreak \pm \vfbreakErr$~Hz, which corresponds to
    $\kperp d_i = \vkdiBreak \pm \vkdiBreakErr$ and $\kperp\rhoi = \vkrhoiBreak \pm \vkrhoiBreakErr$.
    Four-spacecraft phase differencing gives
    $\mathbf{k}\cdot\mathbf{V}/2\pi f = \vkdotV$, supporting the Doppler
    conversion at approximately the 20~\% level around the break. Separately,
    the field remains more closely frozen to the electrons than to the ions
    near $\kperp d_i = 1$. At $\beta_i = 14.4$ the two candidate scales differ by a
    factor of \vrhoiOverdi, which is what makes the distinction measurable
    here. Because the conversion is approximate and only one interval is
    analyzed, we do not claim a definitive identification of the transition
    scale.

    \item {}
    The kinetic-range spectrum constrains the cascade but not the
    dissipation mechanism. The kinetic-range index is
    $\valphaKin \pm \valphaKinStat\,(\mathrm{stat}) \pm \valphaKinWin\,(\mathrm{fit\ window})$, which
    lies closer to the weakly damped dispersive prediction of $-8/3$ than to
    the strongly Landau-damped limiting prediction of $-11/3$
    \citep{boldyrev12,passot15,alexandrova12}, and therefore disfavors the
    strongly damped limiting description for this interval. The kinetic range
    is curved rather than a single power law, so
    any single index is a summary rather than a scaling law. Landau damping,
    transit-time damping and stochastic heating all steepen a cascade and
    cannot be separated by a spectral index; identifying the dissipation
    channel requires velocity-space measurements.
\end{enumerate}

{}
The fluid-scale index from the same broken power-law fit is
$\alpha_1 = \valphaFluid \pm \valphaFluidErr$. As in Section~\ref{sec:tworange}, it
describes the spectrum below the break and is not a scaling law.
The methodology is straightforward to apply to other MMS magnetosheath
intervals and to future missions providing high-cadence burst-mode
electromagnetic measurements.

\section*{Declaration of competing interest}

The authors declare that they have no known competing financial interests or
personal relationships that could have appeared to influence the work reported
in this paper.

\section*{Acknowledgments}

MMS data are publicly available from the NASA Coordinated Data
Analysis Web (CDAWeb) database at \url{https://cdaweb.gsfc.nasa.gov/}.
Data analysis was performed using the PySPEDAS library
\citep{angelopoulos19,grimes22}.

\bibliographystyle{cas-model2-names}
\bibliography{refs-R2}

\end{document}